\documentclass[aps,prb,twocolumn,superscriptaddress,showpacs]{revtex4}
\usepackage{graphicx}
\usepackage{mathrsfs}
\usepackage{amsmath}
\usepackage{subfigure}
\usepackage{bm}
\usepackage{verbatim}
\begin{document}
\title{Unbalanced edge modes and topological phase transition in gated trilayer graphene}
\author{Xiao Li}\email{lixiao@physics.utexas.edu}
\affiliation{Department of Physics, The University of Texas at
Austin, Austin, Texas 78712, USA}
\author{Zhenhua Qiao}\email{zhqiao@physics.utexas.edu}
\affiliation{Department of Physics, The University of Texas at
Austin, Austin, Texas 78712, USA}
\author{Jeil Jung}
\affiliation{Department of Physics, The University of Texas at
Austin, Austin, Texas 78712, USA}
\author{Qian Niu}
\affiliation{Department of Physics, The University of Texas at
Austin, Austin, Texas 78712, USA}\affiliation{International Center for Quantum Materials, Peking University, Beijing 100871, China}

\date{\today}

\begin{abstract}
Gapless edge modes hosted by chirally-stacked trilayer graphene display unique
features when a bulk gap is opened by applying an interlayer potential difference.
We show that trilayer graphene with half-integer valley Hall conductivity
leads to unbalanced edge modes at opposite zigzag boundaries,
resulting in a natural valley current polarizer.
This unusual characteristic is preserved in the presence of Rashba spin-orbit coupling
that turns a gated trilayer graphene into a ${Z}_2$ topological insulator with an odd
number of helical edge mode pairs.
\end{abstract}

\maketitle

\emph{Introduction---}
Gapless edge modes in two-dimensional condensed matter physics
often appear at surfaces or boundaries of materials with
a well-defined topological order in the bulk.
Well-known examples include the chiral quantum Hall edge states
characterized by the first Chern number $\mathcal{C}$~\cite{QHE},
spin-helical edge modes in the
$Z_2$ topological insulators protected by time reversal invariance~\cite{TIreview},
valley-helical edge modes in graphene~\cite{QVHE}, and kink states arising between
regions of inverted bulk orbital moments or valley-Hall
conductivities \cite{volovik,morpurgo,yaowang,vhall_multilayer,kink_highways,kink_BNC,killi}.
Recently, multilayer graphene with chiral stacking has generated lots of interest in the community~\cite{Trilayer}.
Such a system can
develop a valley-Hall conductivity in the presence of
a gap opening mass term \cite{vhall_multilayer},
including gapped bilayer \cite{morpurgo} and single layer \cite{yaowang}.
For all these known examples so far it is expected that
the edge modes (chiral or helical) are distributed in equal numbers at opposite boundaries.

In this Rapid Communication, we study the special behavior of edge modes in chirally-stacked trilayer graphene in the presence of an interlayer potential difference.
We show that gated trilayer graphene has an unusual spatially unbalanced distribution of valley-Hall edge modes
at sample boundaries with unequal pairs of edge modes,
resulting in a natural valley current polarizer.
We also show that the gated trilayer can be brought into a topological insulator phase by
introducing the Rashba spin-orbit coupling (SOC).
In this case the edge states are also distributed unevenly
and the system is characterized by different \emph{odd} numbers of edge modes pairs
located at opposite sample boundaries.

\emph{System Hamiltonian---}
In the following, we present the general form of
the tight-binding Hamiltonian of a gated trilayer graphene
in the presence of Rashba SOC
\begin{eqnarray}
H&=&H_{\text{SLG}}^{\text{T}}+H_{\text{SLG}}^{\text{M}}+H_{\text{SLG}}^{\text{B}}
+t_{\perp}\sum_{i\in \text{T}, j\in \text{M}} c^{\dagger}_{i}c_{j}\notag\\
&+&t_{\perp}\sum_{i\in \text{M}, j\in \text{B}} c^{\dagger}_{i}c_{j}
+U\sum_{i\in \text{T}} c^{\dagger}_{i}c_{i}-U\sum_{i\in \text{B},} c^{\dagger}_{i}c_{i},
\end{eqnarray}
where $H_{\text{SLG}}^{\text{T},\text{M},\text{B}}$ represent respectively
the monolayer graphene Hamiltonian of the top (T), middle (M) and bottom (B) layers, and can be written as
\begin{eqnarray}
H_{\text{SLG}} = -t\sum_{\langle ij\rangle}c^{\dagger}_{i}c_{j}+it_{\text{R}}\sum_{\langle ij\rangle \alpha\beta}(\bm{s}_{\alpha\beta}\times \bm{d}_{ij})_z c^{\dagger}_{i\alpha}c_{j\beta},
\end{eqnarray}
where
$c^{\dagger}_{i}$ creates an electron on site $i$, and $t$
is the intralayer hopping energy between nearest neighbours.
The Rashba SOC \cite{Kane} strength is measured by $t_{\text{R}}$,
where $\bm{s}$ are spin Pauli matrices, and $\bm{d}_{ij}$ describes
a lattice vector pointing from site $j$ to site $i$.
The interlayer hopping $t_{\perp}$
couples two neighbouring layers in a Bernal stacking pattern.
Finally, the gate bias $2U$ is applied by setting the lattice site potentials
to be $+U$, 0, and $-U$ on the top, middle, and bottom layers, respectively.

\emph{Quantum valley-Hall edge modes---}
When a perpendicular electric field is applied on a multilayer graphene,
a nontrivial bulk gap opens to host a quantum valley-Hall state. This is characterized by a valley
Hall conductivity given by ${\sigma}^{v}_{xy} = ({\sigma}^{\text{K}}_{xy}-{\sigma}^{\text{K'}}_{xy})e^2/2h $ for the spinless case, % N \tau_{z} {\rm sgn}(U)/2 e^2/h where $\tau_z = \pm 1$ signal the valley indices of $K$ and $K'$
\cite{vhall_xiao,vhall_multilayer}, where ${\sigma}^{\text{K,K'}}_{xy}$ is obtained by using a continuum model at K or K'. When these two valleys are separated and intervalley scattering is avoided, such valley Hall conductivity assumes an integer (semi-integer) value for even (odd)-$N$ layers of graphene stacks \cite{weakTI}. Such a bulk quantization only has edge correspondence at specific system boundaries. For example, zigzag ribbon geometries with large momentum separation between valleys \cite{brey}
can support valley-Hall edge modes, manifesting the quantized valley-Hall conductivity of the bulk \cite{vhall_multilayer}.
For \emph{even} $N$, we can see an integer number of valley-Hall edge
mode pairs located at both edges in keeping with the integer quantization of valley-Hall conductivity.
For \emph{odd} $N$, however, a qualitatively distinct feature is observed in the valley-Hall edge modes
that we discuss at length in the following. Due to the requirement of absence of intervalley scattering, quantum valley Hall state is considered as a ``weak" topological state, compared to the topologically protected quantum-Hall effect. This scenario resembles the requirement of time-reversal symmetry protection in $Z_2$ topological insulator.

\begin{figure}
\includegraphics[width=8.5cm,angle=0]{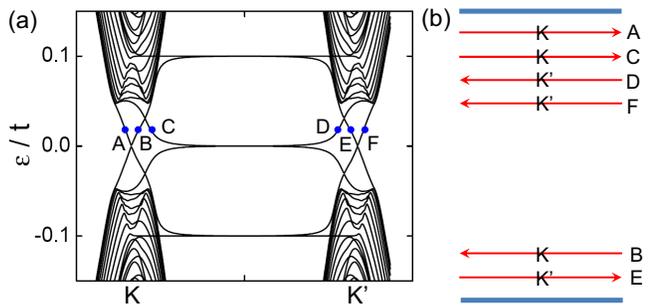}
\caption{(Colour online) (a) Band structure of a zigzag-terminated
trilayer graphene ribbon in presence of an interlayer potential
difference proportional to $U=0.1t$. Letters `A-F' label the six edge states inside the bulk gap with the same
Fermi energy. (b) Edge modes of the edge states labelled in (a).
Spins are doubly degenerate and valleys are associated with the edge states.
Note that there are two pairs of valley-helical edge states at the upper boundary,
while only one pair is localized at the lower boundary.} \label{Fig1}
\end{figure}

Figure~\ref{Fig1}(a) plots the band structure of a zigzag-edged
trilayer graphene ribbon in the presence of an interlayer potential difference $U=0.1t$. One can clearly observe the valley-Hall edge states in the vicinity of each valley inside the bulk gap. At charge neutrality, the edge modes in the system are
formed by the left- and right-going states within the same valley (A, B and E, F)
located at opposite edges as illustrated schematically in Fig.~\ref{Fig1}(b).
As soon as the Fermi level is shifted from neutrality,
the system acquires two additional edge modes,
labelled for electron-doped case with letters C, D in Fig.~\ref{Fig1}(a).
This additional pair of edge modes forms two counter-propagating channels with opposite valley flavours,
and is located at the same boundary of the ribbon,
giving rise to a net valley-polarized current.
The real-space edge location can be reversed by changing
the direction of the external electric field.
At the same time edge asymmetry of electron-hole wave functions
allows for the control of the spatial distribution of edge modes through carrier doping.

Here, we give an intuitive picture on how the valley-Hall edge modes emerge in trilayer graphene. In the absence of interlayer coupling, a gated trilayer graphene is composed of three single layer graphene stacks with their electronic bands relatively shifted by the external potential difference. When the interlayer coupling is further included, gaps open at the bulk band crossing points. Simultaneously, a pair of edge mode is formed at each valley point, i.e., bands labeled by A,B at valley K and E,F at valley K'. Based on the fact that A,F (B,E) are located at the same boundary, we can attribute that the band labeling with A,F(B,E) originates from one of the top (bottom) flat band. And the central flat bands connect with the bulk conduction (valence) bands to form the special band labeled with C,D, which gives rise to the unbalance edge mode.

Therefore, we find an interesting scenario where both the valley polarized current directions
as well as their location at a given edge can be appropriately switched through
electric gating or controlling the carrier density.
The above anomalous features of the unbalanced edge modes in ABC trilayer graphene are preserved when the system is brought into a topological insulator phase.

\emph{$Z_2$ topological insulator edge modes---}
Spin-orbit coupling strengths for carbon atoms in graphene
were estimated to have an extremely small value of the order of $10^{-7}$ eV,
rendering an almost negligible effect for both intrinsic and Rashba SOC in
graphene under realistic conditions \cite{yao,min}.
However, interactions with substrates \cite{RashbaSOC} or adatoms \cite{qiao}
that introduce an additional inter-atomic effective electric field
can increase the Rashba-type SOC to energy scale orders of meV.
Interplay of Rashba SOC with layer inversion symmetry breaking potential in
bilayer graphene was shown to trigger an interesting phase transition
from a quantum valley-Hall phase into a valley-protected topological insulator phase \cite{qiao2}.
We show that a similar topological phase transition is also found in gated trilayer graphene,
but with additional novel features.
\begin{figure}
\centering
\includegraphics[width=8.5cm,angle=0]{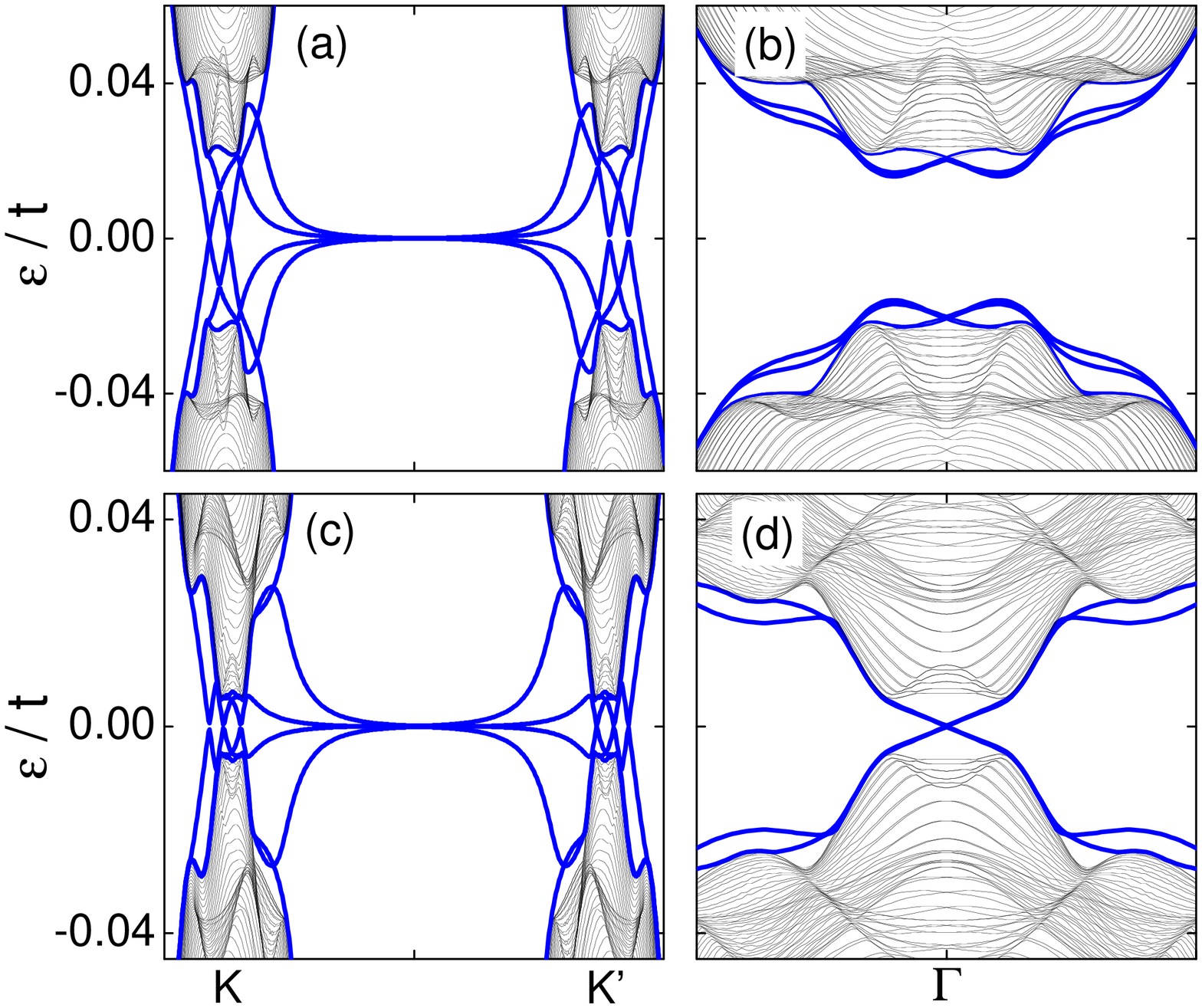}
\caption{(Colour online) Band structures of zigzag [(a) and (c)]
and armchair [(b) and (d)] trilayer graphene ribbons at a fixed
interlayer potential difference $U=0.1t$ with different Rashba SOC.
(a)-(b): $t_R=0.05t$, six edge states, shown in blue (thick) lines, appear at each valley
in zigzag ribbon; no gapless edge state exists inside the bulk gap of armchair ribbon,
though there are some emerging gapped edge bands.
(c)-(d): $t_R=0.12t$, one more pair of edge states are induced at each
valley of zigzag ribbon, and gapless edge states are now formed inside the
bulk gap of armchair ribbon.} \label{Fig2}
\end{figure}
The effect of a Rashba SOC is to split the spin degeneracy
of the bands and lead to an eventual formation of time-reversal invariance protected gapless
edge modes for sufficiently strong Rashba SOC $t_R$. In Fig.~\ref{Fig2}, we show the band structure evolution as a function of $t_R$
for both zigzag [panels (a) and (c)] and armchair [panels (b) and (d)] trilayer
ribbons at a fixed potential difference $U=0.1t$. Edge states are
plotted in blue (thick) lines to distinguish from the bulk bands in black (thin) lines.
\begin{figure}
\centering
\includegraphics[width=8.5cm,angle=0]{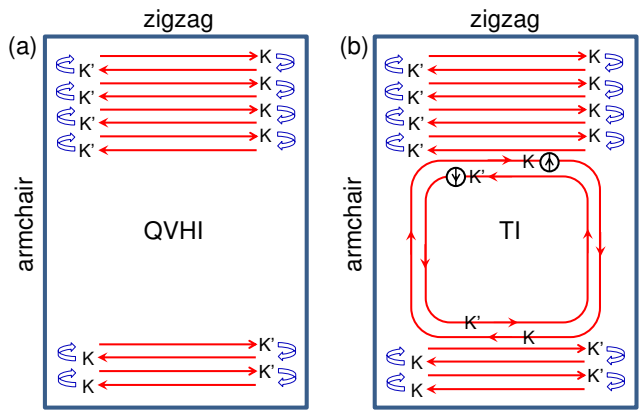}
\caption{(Colour online) Schematic plot of the edge modes corresponding
to the conventional QVHI phase and strong TI phases.
(a) Four (two) pairs of valley-helical edge states are localized at the
upper (lower) boundary. When they encounter the armchair edge
the inter-valley scattering backscatters the edge state associated with valley $K$ to the
counter-propagating edge state encoded with valley $K'$.
(b) The major difference from panel (a) is that one more pair of time-reversal invariance protected
spin-helical edge states emerges at each boundary of the zigzag or armchair trilayer graphene ribbon. Note that in the zigzag ribbon geometry, all the edge modes are associated with both spin and valley degrees of freedom.}
\label{Fig3}
\end{figure}

When a small Rashba SOC is introduced, i.e., $t_R=0.05t$, the bulk band gap of the system starts to decrease
and the spin degeneracy of the bulk bands is lifted.
In Fig.~\ref{Fig2}(a) for the zigzag ribbon, one can observe that three spin-degenerate
pairs of edge states are split into six pairs of non-degenerate edge modes, still preserving the uneven
spatial distribution at opposite edges: four pairs of edge states propagating along one edge,
while only two pairs travelling along the other edge, as shown in Fig.~\ref{Fig3}(a).
Simultaneously, edge modes start to emerge within the bulk gap for the armchair ribbon
as shown in Fig.~\ref{Fig2}(b).

With further increase of Rashba SOC, the bulk band gap continues to decrease while the system remains in the quantum valley-Hall insulator (QVHI) phase.
The system finally reaches a critical point at $t_{\text{R}} = 0.094t$,
where the bulk gap completely closes.
Beyond this point, the bulk band gap reopens, which suggests a topological phase transition.
In the following, we demonstrate the novel characteristics of the resulting edge modes in zigzag-terminated trilayer graphene ribbon after the phase transition.

Figures~\ref{Fig2}(c) and \ref{Fig2}(d) show the band structures for a larger Rashba SOC $t_R=0.12t$.
In zigzag-terminated trilayer graphene, we find a different behaviour with respect to the bilayer case, where one pair of edge states merge and disappear into the bulk bands \cite{qiao2}.
Here, a new pair of edge states emerges from the bulk at both valleys $K$ and $K'$, giving rise to a total of eight pairs of edge
modes inside the bulk gap [see Fig.~\ref{Fig2}(c)].
The resulting edge modes have a rather surprising spatial arrangement
as we show in the schematic plot in Fig.~\ref{Fig3}(b):
five pairs of edge states are located at the upper boundary,
while three pairs are located at the lower boundary.
The odd pairs of spin-helical edge states propagating in a time-reversal invariant
system indicate a topological insulator state.
In this way, we provide an intuitional picture showing that unequal numbers of edge state pairs at opposite sample boundaries can exist in a graphene-based TI system due to the protection of large valley separation.
%We shall show that although valley degree of freedom is still encoded with the edge modes, the valley Chern number is no longer valid in this TI phase.
\begin{figure}
\centering
\includegraphics[width=8.5cm,angle=0]{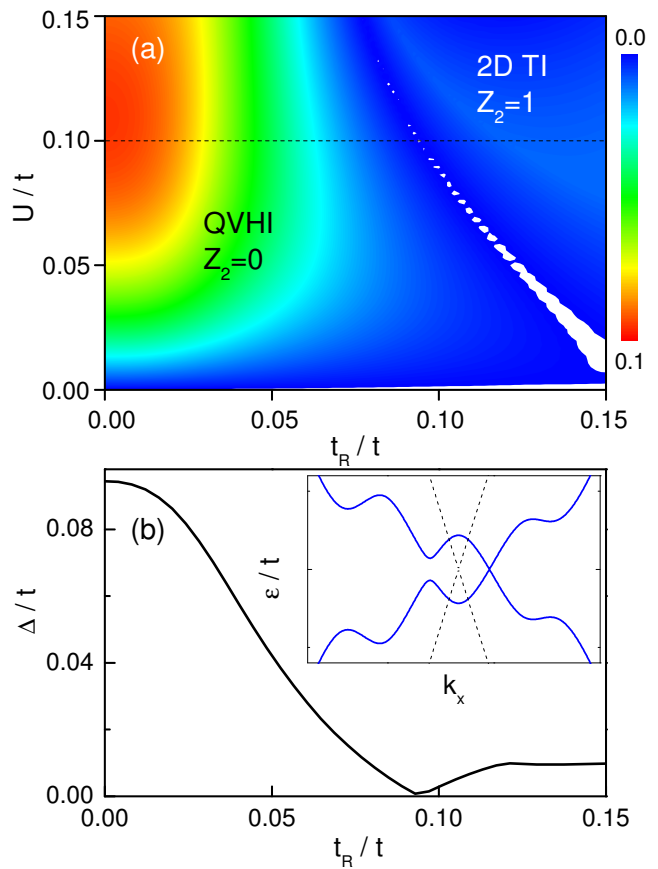}
\caption{(Colour online)
(a) Phase-diagram of ABC trilayer graphene in the parameter space of $t_R$ and $U$. Colors represent the bulk gap size. Two phases are clearly separated: conventional QVHI phase with $Z_2=0$, and the two-dimensional (2D) TI phase with $Z_2=1$. The dashed line trace corresponds to the potential difference $U=0.1t$, which is exhibited in panel (b). The bulk gap shown in panel (b) closes and reopens once in the chosen scale, signaling a topological phase transition from QVHI phase to 2D TI phase. Inset shows that the gap closing point is not at the exact $K$ point. The solid (blue) curves are the bulk band structure of gated trilayer graphene around valley $K$ at the critical Rashba SOC $t^c_R=0.094t$; while the dashed lines plot the band structure of single layer pristine graphene, where the crossing point is exactly at the $K$ point.} \label{Fig5}
\end{figure}

A further confirmation of the topological insulator is the appearance of two pairs of gapless edge
states in the armchair-terminated trilayer graphene ribbon: there is one (odd) pair of edge states flowing along each boundary of the armchair ribbon [see the vertical direction of Fig.~\ref{Fig2}(d)]. Based on the above analysis, we can obtain a schematic diagram of the topological insulator edge
states in trilayer graphene as illustrated in Fig.~\ref{Fig3}(b): there is one pair of time-reversal invariance
protected edge states circulating along any boundaries,
while the remaining edge modes can only propagate along zigzag boundaries due to
the inter-valley scattering present in an armchair edge. Note that the five (three) pairs of edge modes at the upper (lower) boundary are equally-weighted and nontrivial.

Another important signature of TI is the $Z_2$ topological number that characterizes the band topology \cite{Kane,Z2number}. Using the method described in Ref.~[\onlinecite{Z2calculation}], we numerically compute the $Z_2$ topological number for our system, and the results show that $Z_2=0$ before the phase transition, while $Z_2=1$ after the phase transition, consistent with our band structure analysis.

\emph{Phase-diagram---}
To give a complete understanding on how the QVHI phase evolves as functions of $t_R$ and $U$,
we present a ``phase-diagram'' of the bulk band gap $\Delta$ in Fig.~\ref{Fig5}(a).
Colour maps are used to indicate the bulk gap magnitude.
The two separate regimes correspond to a conventional QVHI phase with $Z_2=0$ and $\sigma^{v}_{xy}=3e^2/h$ (considering both spins)
and a topological insulator phase with $Z_2=1$, respectively. As a guide to the eye, we plot the bulk gap dependence
as a function of $t_R$ at a fixed potential difference $U=0.1t$ in Fig.~\ref{Fig5}(b)
that shows one gap closure and reopening as we discussed before.

Different from the topological phase transition in bilayer graphene occurring exactly at $K$ and $K'$ points, in trilayer graphene the bulk band gaps close at some points away from the exact valley $K/K'$ points [see the Inset of Fig.~\ref{Fig5}(b)]. As a consequence, the low-energy continuum model can not correctly capture this nontrivial topological insulator phase. Therefore, although the obtained topological insulator phase in trilayer graphene is still associated with valley degrees of freedom, one can not calculate a well-defined valley-Hall conductivity from the corresponding continuum model.

\emph{Summary and discussions---}
We have studied the spatial imbalance and valley current
polarization of the edge modes in gated trilayer graphene.
The spatially uneven distribution of edge modes
arises from the half-integer quantum valley-Hall conductivity
of trilayer graphene that challenges the conventional understanding
of how edge modes are related with the bulk topology.
These features can in principle be explored in
a doubly gated trilayer device that would allow for a direct control of the edge modes,
either by modifying the carrier doping or reversing the sign of the interlayer potential difference.
Although perfect zigzag trilayer graphene may not be experimentally accessible in current conditions,
it is argued that a substantial contribution of quantum transport in
realistic samples of bulk gapped multilayers might still be mediated by valley-Hall edge modes \cite{jian}.
In such cases, the experimental signatures of edge mode imbalance in
trilayer graphene discussed in this Rapid Communication should have
measurable consequences in electron transport experiments, e.g. in the form of orbital moments \cite{vhall_xiao}
generated by the imbalanced current carrying edge modes.

Another noteworthy finding reported in the present work is that the trilayer graphene can be turned into a topological insulator phase in the presence of sufficiently large Rashba SOC, where the distribution of the edge modes at opposite boundaries remains uneven. Due to the special structure of zigzag ribbon without intervalley scattering, we show that the numbers of edge mode pairs at both boundaries are odd, and most importantly, they are unequal, i.e. five pairs at one boundary while three pairs at the other. When these edge states encounter the armchair edge, only one pair of topologically protected edge modes can survive due to the presence of strong inter-valley scattering.

Finally, we discuss the stability of the unbalanced edge states in the presence of external disorders. As demonstrated in Ref.~\cite{StabilityOfEdgeStates2}, the valley-Hall edge modes can be easily destroyed by short range disorders due to the back scattering between valleys. On the contrary, it is found that long range (smooth) disorders can strongly suppress the back scattering. Moreover, in graphene the impurity scattering in graphene mainly arises from the long-range Coulomb scatterers. Therefore, the proposed unbalanced edge modes are robust against the smooth disorders and should be detectable in a realistic zigzag trilayer graphene ribbon. Recently, it is reported~\cite{WeakTI1,WeakTI2} that surface states in weak topological insulators having even number of Dirac cones are very robust as long as the perturbation does not break the time reversal symmetry.

\emph{Acknowledgements---}
We acknowledge financial support by the NSF (Grant No.~DMR 0906025),
the Welch Foundation (Grant No.~F-1255, TBF1473), NRI-SWAN,
the DOE (Grant No.~DE-FG03-02ER45958, Division of Materials Science and Engineering)
and the MOST Project of China (2012CB921300).

\end{document}